# Room-Temperature Quantum-Confined Stark Effect in Atomically Thin Semiconductor

Michael Engel[1], Mathias Steiner[1,2*]

[1] *IBM Research, Rio de Janeiro, RJ 22290-240, Brazil*

[2] *IBM TJ Watson Research Center, NY 10598, USA*

[*]*msteine@us.ibm.com, mathiast@br.ibm.com*

## ABSTRACT

Electric field-controlled, two-dimensional (2D) exciton dynamics in transition metal dichalcogenide monolayers is a current research focus in condensed matter physics. We have experimentally investigated the spectral and temporal properties of the A-exciton in a molybdenum diselenide ($MoSe_2$) monolayer under controlled variation of a vertical, electric dc field at room temperature. By using steady-state and time-resolved photoluminescence spectroscopies, we have observed dc field-induced spectral shifts and linewidth broadenings that are consistent with the shortening of the exciton's non-radiative lifetime due to field-induced dissociation. We discuss the implications of the results for future developments in nanoscale metrology and exploratory, optoelectronics technologies based on layered, 2D semiconductors.

Layered, atomically thin semiconductors provide unique opportunities for exploring the fundamentals of condensed matter physics. Zero-gap semiconductor graphene, for example, is a proven test system for two-dimensional (2D) charge carrier dynamics in which ballistic transport and quantum Hall effect occur at room temperature[1]. Similarly, monolayers of transition metal dichalcogenides (TMDs)[2] reveal the 2D dynamics of excitons, bound electron-hole pairs that determine a semiconductor's optoelectronic properties[3]. The dynamics of strongly confined excitons in electrostatic fields have been researched since the 1980s when the Quantum-Confined Stark Effect (QCSE) was discovered[4]. Principal features ensure QCSE, electric-field controlled light absorption and emission of excitons confined in semiconductor quantum wells[5-7], are widely exploited in optoelectronic technologies today. In this paper, we report the experimental observation of room-temperature QCSE and electric field-induced dissociation of 2D excitons in monolayer molybdenum diselenide, $MoSe_2$. The discovery of electric field-controlled exciton dynamics in the limit of ultimate intra-layer confinement and high binding energies prevalent in TMD monolayers paves the way for nanoscale metrology and optoelectronic technology based on layered, 2D semiconductors.

In atomically thin semiconductors, the Coulomb interaction between optically excited electrons and holes leads to the formation of excitons[8], an elementary excited state of condensed matter. While multi-layer hetero-structures made of semiconductor thin films have been demonstrated to create nanometer scale quantum wells for 2D confinement of excitons[5-7], a TMD monolayer confines the A-exciton, the energetically lowest-lying bound state, to a spatial domain with a thickness of even less than one nanometer, inside the crystal layer[9], as visualized in Figure 1a. Very high binding energies of the order of 0.5eV[10-12] ensure that A-excitons remain stable and energetically resolved at room temperature[13]. The co-occurrence of efficient intra-layer confinement, high binding energies, and appreciable light absorption and emission has enabled pioneering measurements of electric field induced, spectral modifications of neutral and charged A-excitons in TMD monolayers [12,14-19], including electric dc field-induced Stark shifts at low temperatures[9,20]. With the occurrence of room temperature QCSE in TMDs, technologically viable optoelectronics based on atomically thin semiconductors would come within reach. However, impediments for its experimental discovery have been overlapping spectral signatures of neutral and charged excitons, as well as the complexity of A exciton decay dynamics at elevated temperatures[21-23]. From a theoretical point of view, the quantum mechanical conception of bound hydrogen provides a suitable approach for modeling the electric field controlled dynamics of bound electron-hole pairs in atomically thin semiconductors[24-26]. Inclusion of the electron-hole Coulomb potential in effective mass representation enables the prediction of key observables such as electric field induced, energetic shifts $\Delta E$, as well as modifications of the exciton's homogeneous, lifetime-limited spectral linewidth $\Gamma_{hom}$ and excited state lifetime $\tau_{exciton}$, respectively. The exciton's linewidth and lifetime are inversely proportional and the lifetime-linewidth product is constant; with a lower bound given by Heisenberg's uncertainty relation.

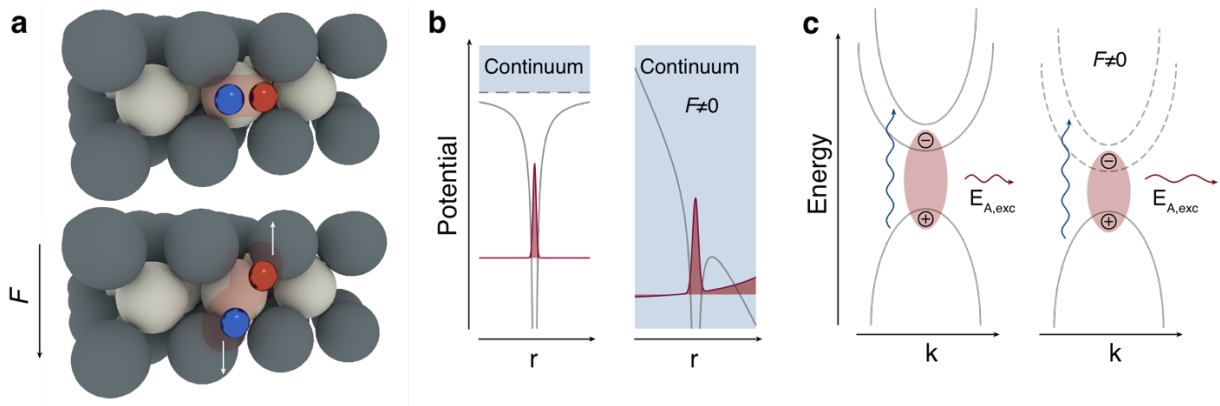

**Figure 1. Quantum confined Stark effect of A excitons in atomically thin semiconductor. a** Artistic rendering of a MoSe$_2$ monolayer (Mo: light gray, Se; dark gray) and a confined exciton (electron: blue; hole: red) in absence (lower panel) and presence (upper panel) of an external, electric dc field $F$ that is oriented perpendicular to the monolayer plane. **b** Schematic potential diagram of the bound exciton state in monolayer MoSe$_2$ in absence (left) and in presence (right) of $F$. The field $F$ causes energetic shifts and broadenings of the exciton and lowers the potential barrier so that the bound state dissociates through carrier tunneling into continuum states. **c** Energy diagram representing the band structure (gray lines) with optical excitation (blue arrow) and relaxation (red arrow) pathways of the A exciton in monolayer MoSe$_2$ in absence (left) and presence (right) of an external, electric dc field $F$.

The experimental observation of electric field controlled A-exciton dynamics in monolayer MoSe$_2$ is schematically rendered in Figure 1. An external, electric dc field $F$ redistributes the wave function of the A exciton, leading to a red shift and broadening of the resonance. By gradually lowering the potential barrier, $F$ increases the probability that the exciton dissociates by means of carrier tunneling into continuum states. Field-induced exciton dissociation constitutes a non-radiative decay channel that effectively lowers the exciton's lifetime as function of $F$. The room temperature occurrence of such electric field-controlled spectral modifications of A excitons at short time scales are a prerequisite for high-speed, optoelectronic modulators and light emitters.

The energy diagram of the MoSe$_2$ monolayer is depicted in Figure 1c. The absorption of a photon is followed by internal conversion into the energetically lower-lying, bound A exciton state. The A exciton decays either through non-radiative relaxation channels or by photon emission, giving rise to a photoluminescence band in the near-infrared spectral range. A low room-temperature photoluminescence quantum yield, $Q \leq 10^{-2}$,[27] indicates that non-radiative exciton decay promoted by phonon scattering[21,28] is the dominant exciton relaxation channel. Due to the complex interrelation of radiative and non-radiative decay contributions at elevated temperatures, A-exciton decay curves are multi-exponential[22,23]. Two principal exciton decay components have been identified: a fast component, of the order of $\tau_f \sim 10^{-13}$ s, associated with purely radiative decay, and a slower decay component, of the order of $\tau_s \sim 10^{-10}$ s, mainly determined by non-radiative decay mechanisms[22,23,29].

An external, electric dc field $F$ should affect both decay components that contribute to $\Gamma_{hom}(F) = \frac{\hbar}{\tau_{exciton}(F)} = \frac{\hbar}{\left(\tau_f^{-1}(F)+\tau_s^{-1}(F)\right)^{-1}} = \hbar \frac{\tau_f(F)+\tau_s(F)}{\tau_f(F) \cdot \tau_s(F)}$ . Although A exciton decay at room temperature occurs predominantly at the slower decay time $\tau_s \gg \tau_f$ [22,23], we expect that the spectral linewidth is limited by the fast decay time, i.e. $\Gamma_{hom}(F) \simeq \hbar/\tau_f(F)$. An additional, inhomogeneous exciton linewidth contribution $\Gamma_{inhom}$, caused by nanoscale heterogeneities and independent of the external field $F$, can be separated off; see Methods Section. Accordingly, we record field induced modifications of the fast decay component $\tau_f(F)$ by means of the exciton's spectral width $\Gamma_{hom}(F)$ while monitoring in $\tau_s(F)$ the onset of field induced exciton dissociation which is predicted to occur at sub-nanosecond time scale[25,26].

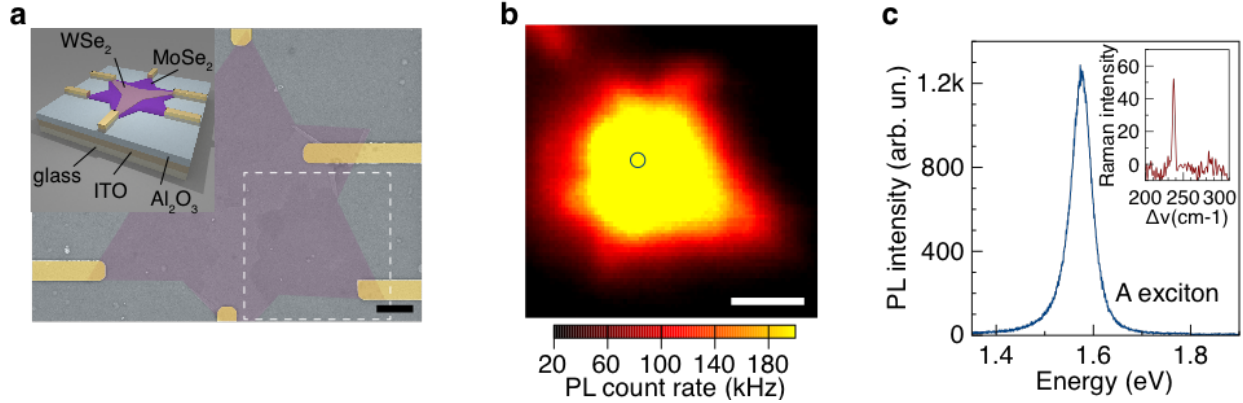

**Figure 2. Monolayer MoSe$_2$ in situ identification and characterization at room temperature. a** Scanning electron microscope image (color enhanced) of a hetero-structure consisting of partially overlapping MoSe$_2$ and WSe$_2$ monolayers. The inset shows a visual representation of the hetero-structure located on top of an optically transparent gate electrode stack with metallic electrodes for application of electric fields. The white dashed line frames the monolayer MoSe$_2$ investigated in this study. The length of the scale bar is 2μm. **b** False color image of the integrated photoluminescence (PL) intensity from monolayer MoSe$_2$ measured at the sample area highlighted by white dashed lines in **a**. The ring indicates the position where photoluminescence spectra and decay profiles were recorded. The length of the scale bar is 2μm. **c** Photoluminescence (PL) spectrum showing the A exciton transition and a Raman spectrum (inset), respectively, of the MoSe$_2$ monolayer.

Figure 2a visualizes the sample investigated in this study. A monolayer of MoSe$_2$ is located at the top of an optically transparent substrate that allows for simultaneous electric control and optical inspection. Details regarding the experimental setup are provided in the Methods Section and reference[30]. The MoSe$_2$ monolayer is partially overlaid by a monolayer of WSe$_2$ forming a star shaped hetero-structure. The experiments have been performed in the pristine MoSe$_2$ monolayer area that is highlighted in Figure 2a. The monolayer area under investigation has a diameter of several microns and exhibits a spatially homogenous photoluminescence response, see Fig. 2b. Figure 2c shows a representative photoluminescence spectrum that exhibits the A-exciton transition at 1.575 eV while a Raman spectrum reveals characteristic vibrational features[10,14]. In addition, atomic force microscope data validating monolayer thickness is provided in Supplemental Figure S1.

For measuring the electric field dependence of the A exciton, a dc voltage, $V$, is applied between an embedded ITO electrode and a metallic electrode at the corner of the $MoSe_2$ monolayer which is electrically isolated from the $MoSe_2$ monolayer by an auxiliary metal oxide layer, see Methods Section. We have performed high-sensitivity measurements to validate the absence of charge carrier transport, $I(V) \leq 10^{-12} A$, for all $V$-values applied in the experiments. Figure 2c shows the photoluminescence spectrum of the A exciton for three representative voltage values while the inset shows the spectrally integrated photoluminescence intensity as function of $V$. Overall, the integrated photoluminescence intensity decreases as $V$ increases, which agrees with experimental findings at low temperature[14] and the theoretical expectation of spectral weight reduction due to an electric field-induced loss of oscillator strength[25]. The spectral area of the A exciton is reduced by up to 40% so that even for the highest voltages applied in our experiments the bound state remains well resolved. An analysis of the electric field distribution reveals that the out-of-plane, or perpendicular, field component is dominant in our experiments, about 4 orders of magnitude larger than the in-plane, or parallel, field component. In the following, for quantifying the effects of the external electric field on the A exciton, we plot the experimental data as function of the perpendicular electric field labeled $F$.

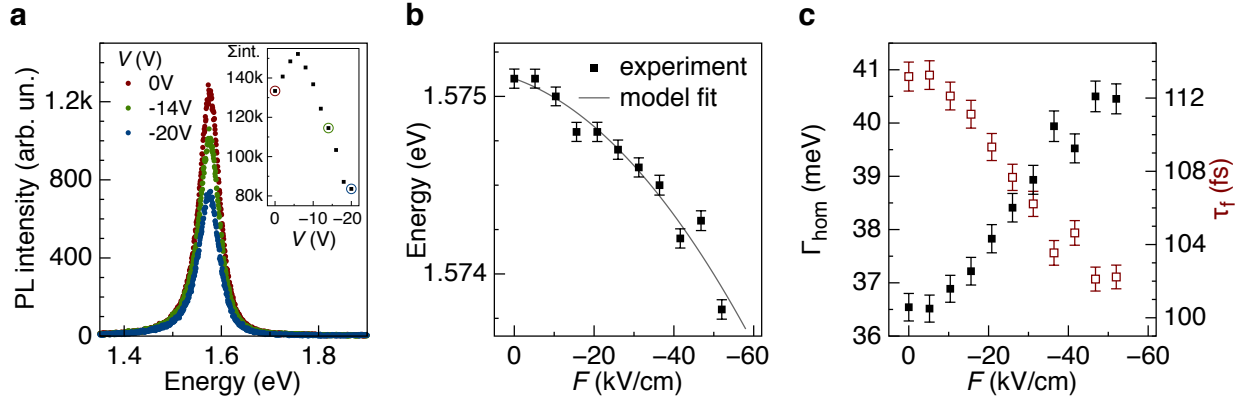

**Figure 3. Electric field induced modulation, shift, and broadening of the A exciton in monolayer MoSe₂. a** Room-temperature photoluminescence (PL) spectra of the A exciton transition in monolayer MoSe₂ for three representative voltage ($V$) values. (Inset) Spectrally integrated photoluminescence intensity of the A exciton as function of $V$. **b** Transition energy of the A exciton as function of electric dc field $F$. **c** Homogeneous spectral width of the A exciton (solid symbols) and lifetime of the fast decay component (open symbols) as function of $F$.

As a key result in Figure 3b, the analysis of the photoluminescence spectrum reveals a dc field-induced, quadratic (Stark) red shift of $\Delta E(52\text{kV/cm}) = -(1.3 \pm 0.1)$meV. By fitting the experimental data with a model function[9,20], see Equation (1) in the Methods Section, we extract the A exciton's permanent dipole moment, $p = (3.4 \pm 0.3)$D, and its polarizability, $\alpha = (0.7 \pm 0.1) \cdot 10^{-6}$Dm/V, similar to values measured in MoS₂ and WSe₂ at low temperature[9,20]. The homogenous, spectral line width contribution of the photoluminescence band as function of the external dc field, $\Gamma_{hom}(F)$, is shown in Figure 3c. The measured zero-field, room-temperature value of about 37mev is in agreement with a previous report[21] and we estimate a fast decay time of $\tau_f \simeq \hbar/\Gamma_{hom} = (102 \pm 1)$fs. Overall, we observe a field-induced, homogeneous linewidth broadening $\Delta\Gamma_{hom}(52\text{kV/cm}) = (3.9 \pm 0.2)$meV which corresponds to a ~10% reduction of the fast decay time, $\Delta\tau_f(52\text{kV/cm}) \simeq \hbar/\Delta\Gamma_{hom}(52\text{kV/cm}) = -(10.9 \pm 0.8)$fs.

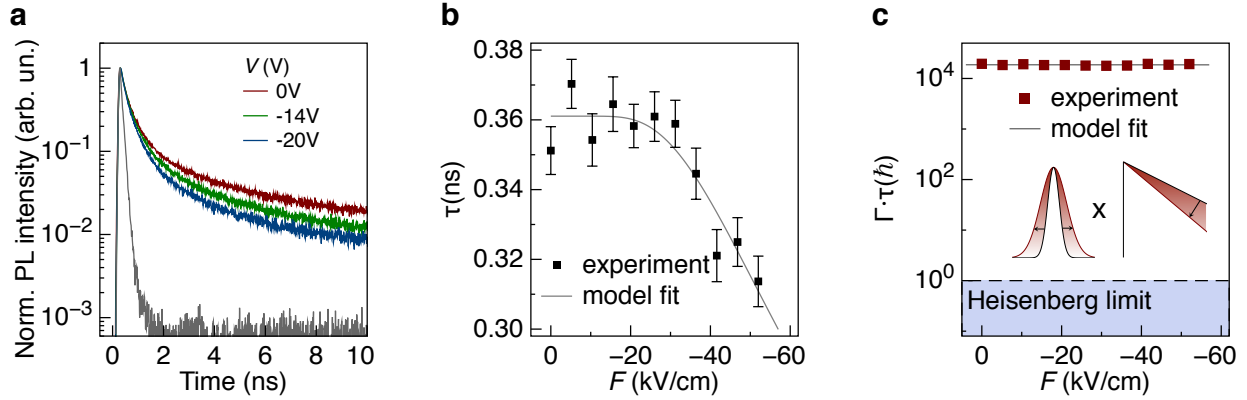

**Figure 4. Electric field induced lifetime shortening of the A exciton in monolayer MoSe₂. a** Room-temperature photoluminescence (PL) decay profiles of the A exciton transition in monolayer MoSe₂ for three representative voltage ($V$)-values. The instrument response function of the experimental setup is shown in gray color. **b** Lifetime $\tau_{1/e}$ of the measured A exciton decay as function of the electric dc field $F$. **c** Product of spectral width $\Gamma_{hom}$ and lifetime $\tau_{1/e}$ of the A exciton as function of $F$.

In Figure 4, we show A-exciton photoluminescence decay as function of the external electric field. Three representative decay curves, taken at the same voltage levels as the spectra in Figure 3a, exhibit gradual exciton lifetime shortenings. A detailed analysis of the A exciton dynamics confirms that the measured photoluminescence decay curves allow us to extract the characteristic time of the slow decay component, $\tau_{1/e} \equiv \tau_s$; see Methods section. In Figure 4b, we plot the 1/e-lifetimes of the A-exciton as function of the electrical dc field. We fit the experimental lifetime-values with a model function that accounts for field-induced exciton dissociation, see Methods section. We observe the threshold for the onset of field-induced exciton dissociation at about $F$=30kV/cm, and, overall, a lifetime shortening $\Delta\tau_s(52\text{kV/cm}) = -(50 \pm 2)\text{ps}$ which amounts to about 14% of the zero field value. Based on the model fit, we extract $F_0 = (138 \pm 5)\text{kV/cm}$ as a characteristic field value for appreciable exciton dissociation in our sample.

Finally, we analyze the relationship between the measured, field-controlled spectral linewidths, see Figure 3c, and exciton lifetimes, see Figure 4b. Figure 4c plots the A-exciton's linewidth-lifetime product $\Gamma_{hom}(F) \cdot \tau_s(F)$; see Equation (5) in the Methods Section. We find that the field-induced increase of spectral linewidth, see Fig. 3c, which is due to the lifetime shortening in the fast decay component, $\Delta\Gamma_{hom}(F) \simeq \hbar/\Delta\tau_f(F)$, is inversely proportional to the measured lifetime reduction, see Fig. 4b, which is due to the onset of exciton dissociation in the slow decay component, $\tau_{1/e}(F) \equiv \Delta\tau_s(F)$. The relationship is constant as function of *F*, however, offset by three orders of magnitude with regards to the lower limit set by Heisenberg's uncertainty relation. At cryogenic temperatures, where the radiative decay component at $\tau_f$ dominates the measured A exciton decay[22,23], $\Gamma(F) \cdot \tau(F)$ should approach the lower bound set by $\hbar$.

The spectroscopic sensitivity of the measured, field-induced effects, $\Delta E(F)/E_0 \simeq 10^{-4}$, $\Delta\Gamma(F)/\Gamma_0 \simeq 10^{-2}$, $\Delta\tau(F)/\tau_0 \simeq 10^{-1}$, demonstrates an experimental accuracy typically unattained in nanoscale materials at room temperature. The unusually high spectral and temporal stability of the luminescence emission, including the robustness of the A-exciton's spectral shape under applied electric fields, is mainly responsible for achieving high signal-to-noise ratios and low error margins in the experimental data. The electric field-controlled spectral and lifetime changes, respectively, demonstrate that precision measurements of 2D exciton dynamics are feasible even in the presence of large spectral broadenings at elevated temperatures. Overall, the demonstrated consistency of energetic shifts, lifetime shortenings, and linewidth broadenings establishes TMD monolayers as a nanoscale model system for the metrology of electric field controlled light matter interaction at room temperature.

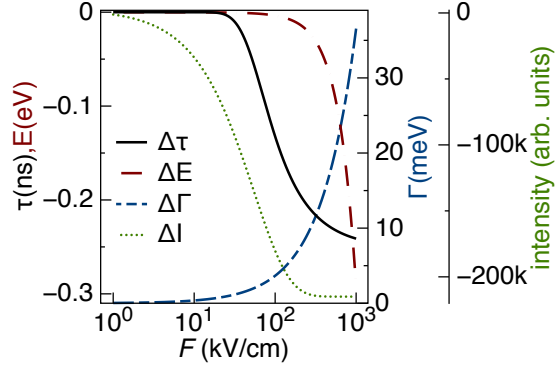

**Figure 5. QCSE metrics for room-temperature optoelectronics based on monolayer MoSe$_2$.** Extrapolation of the energetic shift $\Delta E$, the spectral broadening $\Delta \Gamma$, the lifetime reduction $\Delta \tau$, and the spectrally integrated photoluminescence intensity drop $\Delta I$ as function of the electric dc field $F$ based on the experimental data.

The implications for optoelectronic devices based on TMD monolayers are as follows. The A exciton in MoSe$_2$ monolayers can be energetically tuned and temporally modulated by moderate, external dc fields without the need for cooling the device to cryogenic temperatures; our time-resolved measurements suggest that high-speed operation at GHz frequencies is feasible, see Supplemental Figure S2. A plot of the key metrics based on our experimental data extrapolated across the range of realistic electric field values is shown in Figure 5 and serves as a guide for estimating optoelectronic device performance at room temperature. We note that field-controlled modifications of A exciton absorption are expected to show similar characteristics. The room-temperature QCSE effect could be exploited in TMD photonics for increasing modulation depth and responsivity, respectively, previously demonstrated in graphene-based modulators[31], cavity emitters[32] and detectors[33]; a domain of intense, ongoing research[34]. Furthermore, the results suggest electric field assisted exciton dissociation in combination with substrate engineering[25,26] as a means for improving the efficiency of carrier separation in TMD-based photodetectors[35]. Moreover, the application of QCSE for tuning the emission properties of 2D excitons in TMD based LEDs[36-38] and quantum LEDs[39] show high potential for future exploration.

In conclusion, we have experimentally demonstrated QCSE of the lowest-energy A-exciton in a MoSe$_2$ monolayer at room temperature. The shortening of the exciton's non-radiative lifetime due to field-induced exciton dissociation, consistent with field-induced spectral shifts and linewidth broadenings, confirms the validity of the 2D exciton framework in TMDs and demonstrate that high-precision measurements of quantum properties in atomically thin semiconductors are feasible at room temperature. Moreover, the results demonstrate that optoelectronics technologies based on layered, 2D semiconductors operating at room temperature are finally coming within reach.

**METHODS**

**Materials and Device Manufacturing**

We show a schematic cross section of sample in Supplemental Figure S3a. We prepare the sample based on ITO-coated glass microscope cover slips having a size of 20x20mm$^2$ (06469-AB, Structure Probe, Inc.). We clean the cover slips by subsequently ultra-sonicating them in acetone and isopropanol before placing them in oxygen plasma for 20min. The cover slips are then coated with 100nm of Al$_2$O$_3$ by atomic layer deposition at 250ºC to form a measurement specific substrate stack. In a separate process, TMD monolayers are grown by chemical vapor deposition on sapphire substrates[40]. For transferring as-grown TMD layers from sapphire onto the measurement specific substrate stack, they are spin-coated with PMMA and kept at 60ºC for 10min before being immersed in a 2M KOH solution at 60ºC. During this process step, the PMMA/TMD bilayer detaches from the sapphire substrate enabling the transfer to the desired location of the measurement specific substrate stack. Upon completion of the transfer, the PMMA layer is dissolved in acetone and rinsed with isopropyl alcohol and de-ionized water. We then manufacture metallic electrodes for applying electric fields to the TMD layers by using electron beam lithography. We spin-coat the sample with a PMMA resist (A4, 950 PMMA, MicroChem Corp.) and keep it at 175ºC for 10min prior to electron beam exposure. After completion of electron beam patterning, we develop the resist at 5ºC in a de-ionized water/isopropyl alcohol mixture for 60s. A titanium layer is deposited under high pressure conditions so that it immediately forms titanium oxide which acts as an insulating layer between the TMD layer and a metal electrode that is subsequently deposited by gold evaporation. Finally, the PMMA is lifted off by placing the sample in acetone at 60 ºC for 1h. In order to confirm the monolayer thickness, we perform atomic force microscospy (DI Dimension 3000, Veeco).

**Optoelectronic Measurements**

For combining optical micro-spectroscopy and electrical measurements, we use the setup sketched in Supplemental Figure S3b. A customized, inverted confocal optical microscope equipped with a feedback-controlled piezo-electric scanning stage (P562.3CD controlled by unit E509.C34, Physik Instrumente PI GmbH & Co. KG) is integrated with system equipped with electrical probes (79-6000-R-03 with 7B probe tips, The Micromanipulator Company) that allows to generate electric fields within the sample by means of an external DC voltage source (6430 Sub-Femtoamp Remote SourceMeter SMU, Keithley Inc.). As excitation light source for the photoluminescence measurements, we use a pulsed laser diode operating at 488nm having pulse width of 30picoseconds and a repetition rate of 80MHz (BDL-SMN488, Becker & Hickl GmbH). The laser light is spatially filtered to select a TEM$_{00}$ Gaussian profile and focused onto the sample with an immersion oil objective (100x, N.A. 1.25, Carl Zeiss AG), providing an excitation spot with a diameter of 250nm. For analysis of the photoluminescence emission, we use a spectrometer (HR550, HORIBA Instruments Inc.) equipped with a 300lines/mm grating blazed at 600nm and liquid nitrogen cooled CCD array (1024x256-BIDD-1LS, HORIBA Instruments Inc.), having a resolution limit of 0.3nm. For the acquisition of photoluminescence decays curves, we detect the laser excited emission with a photon counting module (PDM-50-CT, Micro Photon Devices S.r.l) through a band pass filter centered at the photoluminescence peak energy having a bandwidth of 100nm. The detector is connected to a time-correlated single photon counting system (SPC-150, Becker & Hickl GmbH) that is synchronized with the excitation laser, providing an instrument response function with a width of 100picoseconds.

**Exciton linewidth and peak energy**

In order to extract homogeneous and inhomogeneous linewidth contributions from the measured photoluminescence spectra, we fit the experimental data with a Voigt model function. The Voigt function is a convolution of a Lorentz profile and a Gaussian profile[41]:

$$V(x;\sigma,\gamma) = G(x;\sigma) \star L(x;\gamma) \text{ with } G(x;\sigma) = \frac{e^{-x^2/(2\sigma^2)}}{\sigma\sqrt{2\pi}} , L(x;\gamma) = \frac{\gamma}{\pi(x^2+\gamma^2)} \quad (1)$$

The Lorentzian captures the homogeneous, the Gaussian the inhomogeneous linewidth contribution in the measured photoluminescence spectrum. Based on computational model fitting using a data analysis software (pro Fit, Quantum Soft), we extract the homogeneous linewidth contribution $\Gamma_{hom} = 2\gamma$ and the inhomogeneous linewidth contribution $\Gamma_{inhom} = 2\sigma\sqrt{2\ln(2)}$. The results of the analysis are plotted in Supplemental Figure S4. With the same analysis approach, we also obtain from the measured spectra the integrated photoluminescence intensity and exciton peak energy as function of the external, electric field $F$.

**Electric field dependence of exciton energy, dipole moment and polarizability**

For extracting the electric field-dependence of the energetic peak position, $E(F)$, of the A exciton, we fit to the measured data the following model function[9,20]:

$$E(F) = E_0 + pF + \frac{\alpha}{2}F^2 \quad (2)$$

Here, $E_0$ represents the room temperature, zero-field peak energy of the A exciton, $p$ is the permanent dipole moment of the A exciton, and $\alpha$ is its perpendicular, out-of-plane polarizability.

**Exciton decay dynamics and exciton lifetimes**

For the purpose of our analysis, $\tau_s$ and $\tau_f$ signify the characteristic fast and slow photoluminescence decay times, respectively, of the experimentally verified decay contributions in MoSe$_2$[22,23]. At elevated temperature, the probability of purely radiative decay in MoSe$_2$ is much reduced, most likely due to temperature dependent exciton-phonon scattering[21,22] that gradually diminishes the relative strength of the fast A exciton decay contribution as function of T[22,23]. The measured room-temperature photoluminescence decay is dominated by an effective, slow decay component characterized by $\tau_s$ which occurs at a time scale that can be fully resolved given the actual instrument response time, see Figure3a. We analyze the measured data with a multi-exponential model function convoluted with the measured instrument response function IRF($t$) using a data analysis software (Decayfit 1.3, FluorTools) where $\rho_i, \tau_i$ characterize amplitude and decay time, respectively, of the i-th decay component:

$$I(t) = \sum_i \rho_i \exp\left\{-\frac{t}{\tau_i}\right\} \star IRF(t) \quad (3)$$

For modeling the photoluminescence decay at larger times t $\gg \tau_s$, the inclusion of a background decay contribution at nanosecond time scale is required. We obtain reasonable fit results using a bi-exponential model function based on Equation (3) with a background decay contribution at $\tau_{bkg} \simeq 3$ns that is found to be independent of the electric dc field. We note that the principal exciton decay time $\tau_s$ extracted from the bi-exponential model fit and the 1/e-lifetimes extracted directly from the measured photoluminescence decay curves are found to be identical within the experimental resolution, see Supplemental Figure S5. By following the interpretation[22] of the fast exciton decay time as the intrinsic, radiative decay time, $\tau_f \equiv \tau_{rad}$, and assuming that the slow decay time represents the non-radiative decay, $\tau_s \equiv \tau_{non-rad}$, we obtain agreement with the field-dependence of the independently measured, integrated photoluminescence intensity, $I(F) \propto k_{rad}/k_{non-rad} = \tau_s/\tau_f$, see Supplemental Figure S6.

**Electric field dependence of exciton lifetimes**

At room temperature, we approach the field-dependence of the measured, excited state lifetime of the A exciton in MoSe$_2$ through the field-dependence of the slow exciton decay component[23] characterized by $\tau_s$:

$$\tau_s(F) = \frac{1}{k_s(F)} = \frac{1}{k_{s,0}+b\cdot exp\left\{-\frac{F_0}{F}\right\}} \quad (4)$$

Here, the associated decay rate constant, $k_{s,0}$, is the inverse of the room temperature, 1/e-lifetime at zero field which is determined by non-radiative decay processes[21]. The term containing the exciton dissociation field, $F_0$, and the fit parameter, b, capture a field-induced, hydrogen-type exciton dissociation process, see e.g.[25], that represents an externally controlled, non-radiative exciton decay channel that occurs at time scale $\tau_s$.

**Electric field dependence of linewidth-lifetime product**

For analyzing the proportionality of experimentally observed, electric field-induced linewidth and lifetime modifications, respectively, we decompose the spectral linewidth into a field-independent term $\Gamma_{hom,0}$ that represents the measured, room temperature full-width-at-half-maximum linewidth at zero field and a field-dependent term $\Delta\Gamma_{hom}(F)$. Furthermore, we represent the exciton lifetime by the measured, slow decay time $\tau_{1/e} \equiv \tau_s \gg \tau_f$. We decompose $\tau_s$ into a field-independent term $\tau_{s,0}$ that represents the measured, room temperature 1/e-lifetime at zero field and a field-dependent term $\Delta\tau_s(F)$. Finally, we plot in Fig. 4c the following relation:

$$\Gamma_{hom}(F) \cdot \tau_s(F) = \left(\Gamma_{hom,0} + \Delta\Gamma_{hom}(F)\right) \cdot \left(\tau_{s,0} + \Delta\tau_s(F)\right) \quad (5)$$

**ACKNOWLEDGEMENT**

The authors thank Xin-Quan Zhang and Yi-Hsien Lee (National Tsing-Hua University, Taiwan) for the provision of the WSe$_2$/MoSe$_2$ monolayer hetero-structure, Damon Farmer (IBM Research) for expert technical assistance, Jaione Tirapu-Azpiroz (IBM Research) for electrostatic field simulations, and Peter W. Bryant (IBM Research) for discussion related to the exciton lifetime-linewidth analysis. Furthermore, we thank Phaedon Avouris, Shu-Jen Han, and Ulisses Mello (IBM Research) for support.

**REFERENCES**

1. Geim, A. K. & Novoselov, K. S. The rise of graphene. *Nat Mater* **6,** 183–191 (2007).
2. Novoselov, K. S. *et al.* Two-dimensional atomic crystals. *Proc Natl Acad Sci USA* **102,** 10451–10453 (2005).
3. Klingshirn, C. F. *Semiconductor Optics; 4th ed.* (Springer, 2012). doi:10.1007/978-3-642-28362-8
4. Miller, D. A. B. *et al.* Band-edge electroabsorption in quantum well structures: the quantum-confined stark effect. *Phys Rev Lett* **53,** 2173–2176 (1984).

5. Miller, D. A. B. *et al.* Electric field dependence of optical absorption near the band gap of quantum-well structures. *Phys Rev B* **32,** 1043–1060 (1985).
6. Polland, H., Schultheis, L., Kuhl, J., Göbel, E. & Tu, C. Lifetime enhancement of two-dimensional excitons by the quantum-confined Stark effect. *Phys Rev Lett* **55,** 2610–2613 (1985).
7. Kash, J. A., Mendez, E. E. & Morkoç, H. Electric field induced decrease of photoluminescence lifetime in GaAs quantum wells. *Appl Phys Lett* **46,** 173–175 (1985).
8. Frenkel, J. On the transformation of light into heat in solids. I. *Phys. Rev.* **37,** 17–44 (1931).
9. Klein, J. *et al.* Stark effect spectroscopy of mono- and few-layer $MoS_2$. *Nano Lett* **16,** 1554–1559 (2016).
10. Ugeda, M. M. *et al.* Giant bandgap renormalization and excitonic effects in a monolayer transition metal dichalcogenide semiconductor. *Nat Mater* **13,** 1091–1095 (2014).
11. Hanbicki, A. T., Currie, M., Kioseoglou, G., Friedman, A. L. & Jonker, B. T. Measurement of high exciton binding energy in the monolayer transition-metal dichalcogenides WS2 and WSe2. *Solid State Commun* **203,** 16–20 (2015).
12. Zhu, B., Chen, X. & Cui, X. Exciton binding energy of monolayer $WS_2$. *Sci Rep* **5,** 4403–5 (2015).
13. Mak, K. F., Lee, C., Hone, J., Shan, J. & Heinz, T. F. Atomically thin $MoS_2$: a new direct-gap semiconductor. *Phys Rev Lett* **105,** 136805 (2010).
14. Ross, J. S. *et al.* Electrical control of neutral and charged excitons in a monolayer semiconductor. *Nat Commun* **4,** 1474–1476 (2013).
15. Mak, K. F. *et al.* Tightly bound trions in monolayer $MoS_2$. *Nat Mater* **12,** 207–211 (2013).
16. Seyler, K. L. *et al.* Electrical control of second-harmonic generation in a $WSe_2$ monolayer transistor. *Nature Nanotech* **10,** 407–411 (2015).
17. Chernikov, A. *et al.* Electrical tuning of exciton binding energies in monolayer $WS_2$. *Phys Rev Lett* **115,** 126802 (2015).
18. Matsuki, K., Pu, J., Kozawa, D. & Matsuda, K. Effects of electrolyte gating on photoluminescence spectra of large-area $WSe_2$ monolayer films. *Japanese Journal of Applied Physics* **55,** 06GB02 (2016).
19. Vella, D. *et al.* Unconventional electroabsorption in monolayer $MoS_2$. *2D Materials* **4,** 021005–10 (2017).
20. Chakraborty, C. *et al.* Quantum-confined Stark effect of individual defects in a van der waals heterostructure. *Nano Lett* **17,** 2253–2258 (2017).
21. Selig, M. *et al.* Excitonic linewidth and coherence lifetime in monolayer transition metal dichalcogenides. *Nat Commun* **7,** 13279 (2016).
22. Robert, C. *et al.* Exciton radiative lifetime in transition metal dichalcogenide monolayers. *Phys Rev B* **93,** 205423 (2016).
23. Godde, T. *et al.* Exciton and trion dynamics in atomically thin $MoSe_2$ and $WSe_2$: effect of localization. *Phys Rev B* **94,** 165301 (2016).
24. Pedersen, T. G. Exciton Stark shift and electroabsorption in monolayer transition-metal dichalcogenides. *Phys Rev B* **94,** 125424 (2016).
25. Scharf, B. *et al.* Excitonic Stark effect in $MoS_2$ monolayers. *Phys Rev B* **94,** 245434 (2016).
26. Haastrup, S., Latini, S., Bolotin, K. & Thygesen, K. S. Stark shift and electric-field-induced dissociation of excitons in monolayer $MoS_2$ and $hBN/MoS_2$ heterostructures. *Phys*

*Rev B* **94,** 041401 (2016).
27. Chen, H. *et al.* Manipulation of photoluminescence of two-dimensional MoSe$_2$ by gold nanoantennas. *Sci Rep* **6,** 22296 (2016).
28. Dey, P. *et al.* Optical coherence in atomic-monolayer transition-metal dichalcogenides limited by electron-phonon interactions. *Phys Rev Lett* **116,** 127402 (2016).
29. Palummo, M., Bernardi, M. & Grossman, J. C. Exciton radiative lifetimes in two-dimensional transition metal dichalcogenides. *Nano Lett* **15,** 2794–2800 (2015).
30. Sundaram, R. S. *et al.* Electroluminescence in single layer MoS$_2$. *Nano Lett* **13,** 1416–1421 (2013).
31. Liu, M. *et al.* A graphene-based broadband optical modulator. *Nature* **474,** 64–67 (2011).
32. Engel, M. *et al.* Light-matter interaction in a microcavity-controlled graphene transistor. *Nat Commun* **3,** 906 (2012).
33. Furchi, M. *et al.* Microcavity-integrated graphene photodetector. *Nano Lett* **12,** 2773–2777 (2012).
34. Sun, Z., Martinez, A. & Wang, F. Optical modulators with 2D layered materials. *Nat Photonics* **10,** 227–238 (2016).
35. Mak, K. F. & Shan, J. Photonics and optoelectronics of 2D semiconductor transition metal dichalcogenides. *Nat Photonics* **10,** 216–226 (2016).
36. Pospischil, A., Furchi, M. M. & Mueller, T. Solar-energy conversion and light emission in an atomic monolayer p–n diode. *Nature Nanotech* **9,** 257–261 (2014).
37. Ross, J. S. *et al.* Electrically tunable excitonic light-emitting diodes based on monolayer WSe2 p–n junctions. *Nature Nanotech* **9,** 268–272 (2014).
38. Baugher, B. W. H., Churchill, H. O. H., Yang, Y. & Jarillo-Herrero, P. Optoelectronic devices based on electrically tunable p–n diodes in a monolayer dichalcogenide. *Nature Nanotech* **9,** 262–267 (2014).
39. Palacios-Berraquero, C. *et al.* Atomically thin quantum light-emitting diodes. *Nat Commun* **7,** 12978 (2016).
40. Zhang, X.-Q., Lin, C.-H., Tseng, Y.-W., Huang, K.-H. & Lee, Y.-H. Synthesis of Lateral Heterostructures of Semiconducting Atomic Layers. *Nano Lett* **15,** 410–415 (2015).
41. Olivero, J. J. & Longbothum, R. L. Empirical fits to the Voigt line width: a brief review. *Journal of Quantitative Spectroscopy and Radiative Transfer* **17,** 233–236 (1977).

# Supplementary Information

# Room-Temperature Quantum-Confined Stark Effect in Atomically Thin Semiconductor

Michael Engel[1], Mathias Steiner[1,2*]

[1] *IBM Research, Rio de Janeiro, RJ 22290-240, Brazil*

[2] *IBM TJ Watson Research Center, NY 10598, USA*

[*]msteine@us.ibm.com, mathiast@br.ibm.com

**Content: Supplemental Figures S1-S6**

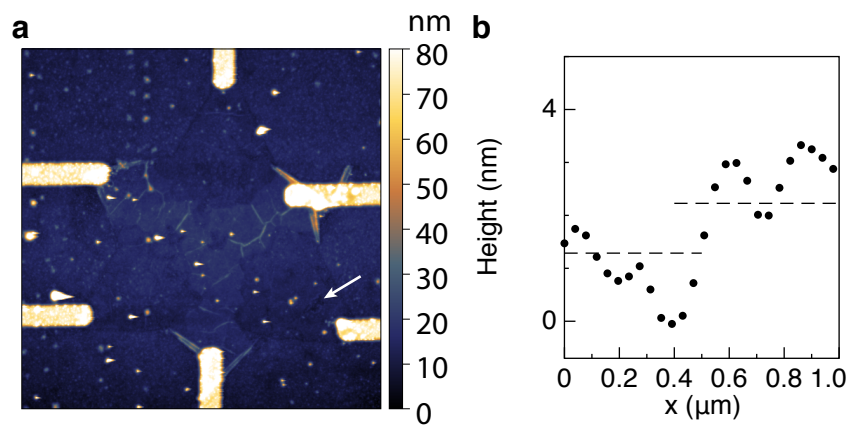

**Supplemental Figure S1. Atomic force microscope measurement of the sample. a** AFM image of the sample discussed in the main manuscript. Size: (20x20) μm². **b** Cross-section taken along the white arrow shown in the image in **a**, at the edge of the monolayer $MoSe_2$ area highlighted in Figure 2 of the main manuscript.

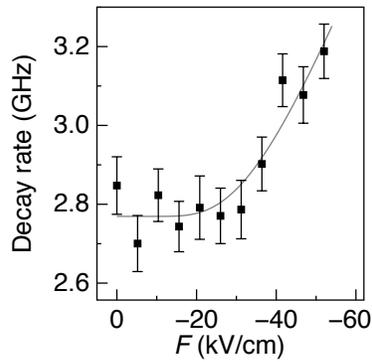

**Supplemental Figure S2. Room-temperature A exciton photoluminescence decay rate in monolayer MoSe$_2$ as function of electric dc field $F$.** The experimental data (squares) and model fit (line) represent the $(\tau_{1/e})^{-1}$-values derived from the measured exciton decay curves; see Figure 3b in the main manuscript.

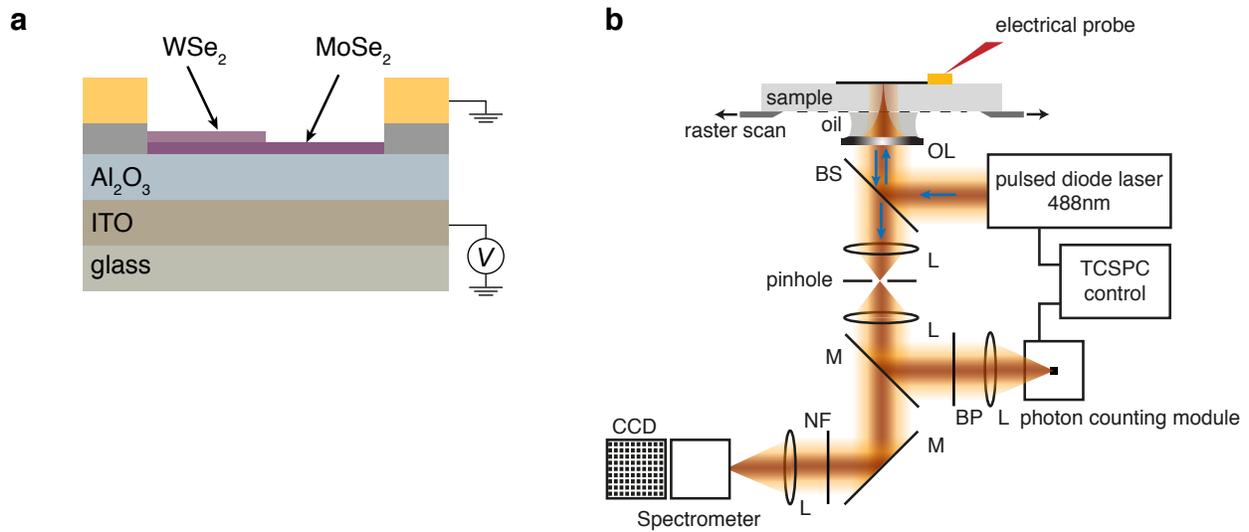

**Supplemental Figure S3. Cross sectional schematic view of the sample and the experimental setup. a** The TMD monolayers are contacted by insulating layer of TiO$_2$ (gray) which is covered by a Au film (orange). The electrical connections for applying a bias voltage $V$ perpendicular to the MoSe$_2$ monolayer are also indicated. **b** Experimental setup for optical micro-spectroscopy with application of electrical fields as discussed in the Method Section.

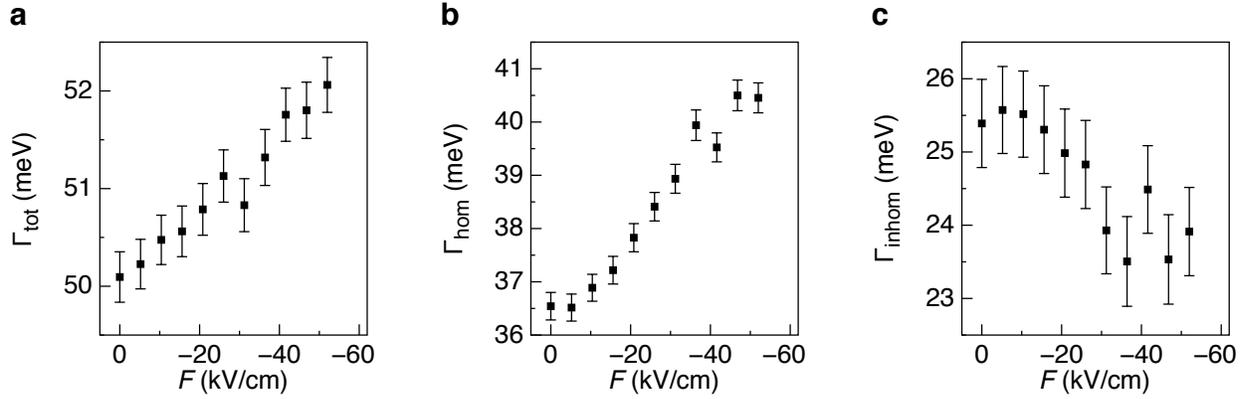

**Supplemental Figure S4. Linewidth analysis of room-temperature A exciton photoluminescence spectrum in monolayer MoSe$_2$. a** Spectral linewidth of the photoluminescence band (full-width-at-half-maximum) as function of the external dc field $F$ based on the Voigt model fit analysis described in the Methods Section. **b** Linewidth of the homogenous contribution as function of $F$ as discussed in the main manuscript. **c** Linewidth of the inhomogeneous contribution as function of $F$.

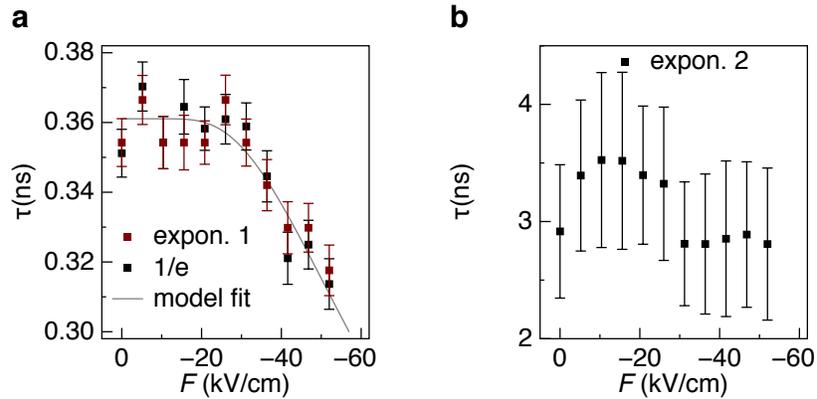

**Supplemental Figure S5. Room-temperature A exciton photoluminescence decay dynamics in monolayer MoSe$_2$ as function of electric dc field $F$. a** 1/e-lifetimes of A exciton decay curves and fitted decay times (labeled expon. 1) based on the analysis procedure outlined in the Methods Section. **b** Characteristic time of the background decay component (labeled expon. 2) extracted from the model fit.

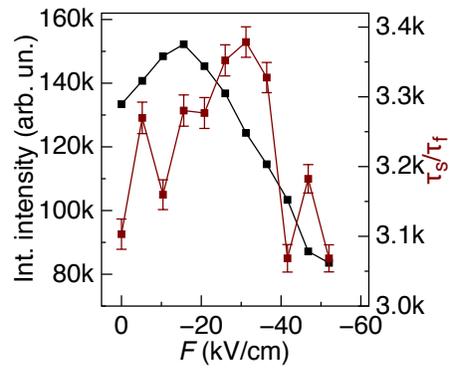

**Supplemental Figure S6. Integrated, room-temperature A exciton photoluminescence intensity of monolayer MoSe$_2$ as function of electric dc field $F$. a** Data taken from photoluminescence spectral analysis, see inset of Figure 2a of the main manuscript. **b** Ratio between slow and fast MoSe$_2$ decay times as function of the electric field $F$.